\documentclass[12pt]{article}

\usepackage{graphics,epsfig}

\begin{document}

\setlength{\hsize}{16cm}
\setlength{\vsize}{20cm}
\renewcommand{\baselinestretch}{1.2}

\title{\bf $SU(3)_c\otimes SU(3)_L\otimes U(1)_X$ as an $E_6$ Subgroup}
\author{{\bf Luis A. S\'anchez} \\
{\it Depto. de F\'\i sica, Universidad Nacional de Colombia,} \\
{\it A.A. 3840,
Medell\'\i n, Colombia }\\ {\bf William A. Ponce} \\ {\it Depto. de F\'\i
sica, Universidad de Antioquia,} \\
{\it A.A. 1226, Medell\'\i n, Colombia} \\
{\bf R. Mart\'\i nez} \\ {\it Depto. de  F\'\i sica, Universidad Nacional de Colombia,} \\
{\it Bogot\'a, Colombia.}}

\date{}
\maketitle

\begin{abstract}

{An extension of the Standard Model to the local gauge group
$SU(3)_c\otimes SU(3)_L\otimes U(1)_X$ which is a subgroup of the
electroweak-strong unification group $E_6$ is analyzed. The mass
scales, the gauge boson masses, and the masses for the spin 1/2
particles in the model are calculated. The mass differences
between the up and down quark sectors, between the quarks and
leptons, and between the charged and neutral leptons  in one
family are explained as a consequence of mixing of ordinary with
exotic fermions implied by the model. By using experimental
results we constrain the
mixing angle between the two neutral currents and the mass of the
additional neutral gauge boson to be $-0.0015\leq\sin\theta\leq
0.0048$ and $M_{Z_2}\geq 600$ GeV at 95$\%$ CL. The
existence of a Dirac neutrino for each family with a mass of the
order of the electroweak mass scale is predicted.}
\end{abstract}


\section{Introduction} The Standard Model (SM) local gauge group
$G_{SM}\equiv SU(3)_c\otimes SU(2)_L\otimes U(1)_Y$ with the
$SU(2)_L\otimes U(1)_Y$ sector hidden\cite{sm} and $SU(3)_c$
confined\cite{su3}, is in excellent agreement with all present
experiments. However, the SM can not explain some issues like
hierarchical masses and mixing angles, charge quantization, CP
violation, etc., and many physicists believe that it does not
represent the final theory, but serves as an effective theory,
originating from a more fundamental one. So, extensions of the SM
are always worthy to be considered.

So far there are not yet theoretical neither experimental facts
which point toward what lies beyond the SM, and the best approach
maybe to depart from it as little as possible. In this regard,
$SU(3)_L\otimes U(1)_X$ as a flavor group has been introduced at
several times in the literature; first as a family independent
theory as in the SM\cite{lee}, and next as a family
structure\cite{fp} which points to a natural explanation of the
total number of families in nature. Several of the models studied
in the literature are inconsistent in the sense that they are
anomalous, vectorlike, or include more than three families with
light neutral particles. Others include spin 1/2 particles with
exotic electric charges\cite{fp}, etc.. In this paper we studied
$SU(3)_L\otimes U(1)_X$ as an $E_6$ subgroup\cite{e6} and do some
phenomenological calculations in order to set the different mass
scales in the model and calculate the masses for all the spin 1/2
particles in one family. In our analysis several see-saw
mechanisms are implemented.

The paper is organized in the following way. In Sec. 2 we
introduce the model as an anomaly free theory, based on the local
gauge group $SU(3)_c\otimes SU(3)_L\otimes U(1)_X$, and show that
it is a subgroup of the electroweak-strong unification group
$E_6$. In Sec. 3 we describe the scalar sector needed to
break the symmetry and to produce masses to all the fermion fields
in the model. In Sec. 4 we analyze the gauge boson sector
paying special attention to the two neutral currents present in the model and
their mixing. In Sec. 5 we analyze the fermion mass
spectrum. In Sec. 6 we use experimental results in order to
constraint the mixing angle between the two neutral currents and
the mass scale of the new neutral gauge boson, and in Sec. 7 we present our conclusions. A technical appendix with
details of the diagonalization of the $5\times 5$ mass matrix for
the spin 1/2 neutral leptons in the model is presented at the end.

\section{The fermion content of the model} 
First let us see what is the fermion content of the model.
\subsection{$SU(3)_c\otimes SU(3)_L\otimes
U(1)_X$ as an anomaly free model} In what follows we assume that
the electroweak gauge group is $SU(3)_L\otimes U(1)_X\supset
SU(2)_L\otimes U(1)_Y$, that the left handed quarks (color
triplets) and left-handed leptons (color singlets) transform as
the $3$ and $\bar{3}$ representations of $SU(3)_L$ respectively,
that $SU(3)_c$ is vectorlike as in the SM and, contrary to the
models in Ref.\cite{fp}, that the several anomalies are cancelled
individually in each family. So, we start with

\[Q_L=\left(\begin{array}{c}u\\d\\D \end{array}\right)_L, \hspace{1cm}
\psi_L=\left(\begin{array}{c}e^-\\ \nu_e\\N_1^0\end{array}\right)_L, \]
\noindent where $D_L$ is an $SU(2)_L$ singlet exotic quark of electric
charge $-1/3$, and $N_{1L}^0$ is also an $SU(2)_L$ singlet exotic lepton
of zero electric charge.

If the \{$SU(3)_c,\; SU(3)_L,\; U(1)_X$\} quantum numbers are
\{$3,3,X_Q$\} for $Q_L$, and \{$1,\bar{3},X_L$\} for $\psi_L$, two more
lepton multiplets with quantum numbers \{1,$\bar{3},X_i\},\; i=1,2$ must
be introduced in order to cancel the $[SU(3)_L]^3$ anomaly; each one of
those multiplets must include one $SU(2)_L$ doublet and one singlet of
exotic leptons. The quarks fields $u^c_L,\; d^c_L$ and $D^c_L$ color
anti-triplets and $SU(3)_L$ singlets must also be introduced in order to
cancel the $[SU(3)_c]^3$ anomaly.  Then the hypercharges ($X_Q,\; X_L,\;
X_1,\;  X_2,$ etc.) must be chosen in order to cancel the anomalies
$[SU(3)_c]^2U(1)_X,\;  [SU(3)_L]^2U(1)_X, \; [grav]^2U(1)_X$ and
$[U(1)_X]^3$, where $[grav]^2U(1)_X$ stands for the gravitational anomaly \cite{wolfe}.

We use for the symmetry-breaking chain $SU(3)_L\longrightarrow SU(2)_L
\otimes U(1)_Z$ the branching rule $3\longrightarrow 2(1/6)+1(-1/3)$,
where the numbers in parenthesis are the $Z$ hypercharge value (which in
turn implies $\bar{3}\longrightarrow 2(-1/6)+1(1/3)$). Then, by using for
the electric charge generator $Q=T_{3L}+Z+X$ where $T_{3L}=\pm 1/2$ is the
$SU(2)_L$ third isospin component, we have that $X_Q=0$, $X_u=-2/3$ and
$X_D=X_d=1/3$. For those values the anomaly $[SU(3)_c]^2U(1)_X$ is
automatically cancelled.

By the same token $X_L=-1/3$.  Then the $[SU(3)_L]^2U(1)_X$ anomaly
cancellation condition reads $X_1+X_2=1/3$ and the $[U(1)_X]^3$ anomaly
cancellation condition reads $X_1^3+X_2^3=7/27$. The solution to this pair
of equations is $X_1=-1/3$ and $X_2=2/3$ (or vice versa because the two
equations are 1,2 symmetric).

We end up with the following anomaly free multiplet structure for
this model:
\[\begin{array}{||c|c|c|c||}\hline\hline
Q_L=\left(\begin{array}{c}u\\d\\D \end{array}\right)_L & u^c_L & d^c_L&
D^c_L \\ \hline (3,3,0) & (\bar{3},1,-{2\over 3}) & (\bar{3},1,{1\over 3})
& (\bar{3},1,{1\over 3}) \\ \hline\hline \end{array} \]

\[\begin{array}{||c|c|c||}\hline\hline
\psi_L=\left(\begin{array}{c}e^-\\ \nu_e\\N^0_1\end{array}\right)_L &
\phi_L=\left(\begin{array}{c} E^- \\ N_2^0\\N^0_3\end{array}\right)_L &
\chi_L=\left(\begin{array}{c} N^0_4\\E^+ \\e^+\end{array}\right)_L \\ \hline
(1,\bar{3},-{1\over 3}) & (1,\bar{3},-{1\over 3}) & (1,\bar{3},{2\over 3})\\
\hline\hline
\end{array} \]

\subsection{$SU(3)_L\otimes SU(3)_c\otimes U(1)_X$ as an $E_6$ subgroup}
Models constructed with $E_6$ as the grand unified theory (GUT) group\cite{e6} utilize
the complex 27 dimensional representation for the left-handed
fermions in one family. The models can be divided into two
classes, depending on whether they emphasize the $SO(10)$ or the
$SU(3)_L\otimes SU(3)_c\otimes SU(3)_R$ subgroups. For models in
the second category the fundamental representation of $E_6$ has
the following branching rule under $[SU(3)]^3\equiv SU(3)_L\otimes
SU(3)_c\otimes SU(3)_R$:

\[27\longrightarrow (3,3,1)\oplus (\bar{3},1,3)\oplus (1,\bar{3},\bar{3})\]
where the particle content of each term is:
\[(3,3,1)=(u,d,D)_L \;\; ; \;
(\bar{3},1,3)=\left(\begin{array}{ccc} N^0 & E^- & e^- \\ E^+ &
N^{0c} & \nu_e \\ e^+ & \nu^c_e & M^0 \end{array}\right)_L  \;\;
;\; (1,\bar{3},\bar{3})=(u^c,d^c,D^c)_L  .\]

When $SU(3)_R$ breaks into $U(1)_a\otimes U(1)_b$ we have the
branching rule $3\longrightarrow (a)(b)+(-a)(b)+(0)(-2b)$ which
shows that taking $b=1/3$ we can identify $U(1)_b=U(1)_X$ and it implies that $(\bar{3},1,3)=\psi_L \oplus\phi_L\oplus\chi_L$. This allows
us to identify $N_{3L}^0=\nu^c_{eL}$ as the right handed neutrino
field (in the weak basis) and $N^0_{4L}= N_{2L}^{0c}$, as far as
we stay in the breaking chain
\[ [SU(3)]^3\longrightarrow SU(3)_L\otimes SU(3)_c\otimes SU(2)_R\otimes
U(1)'\longrightarrow SU(3)_L\otimes SU(3)_c\otimes U(1)_X\]

Let us mention that the lower dimensional $E_6$ irreducible
representations (irreps) are 1, 27, and 78, where irreps 1 and 78
are real representations and irrep 27 is complex and
chiral\cite{e6}. In the electroweak-strong unification theory based
on $E_6$ the fermion fields are placed in irrep 27 and the gauge
fields in irrep 78.

\section{The scalar sector}
Our aim is to break the symmetry in the way:
\[SU(3)_c\otimes SU(3)_L\otimes U(1)_X\longrightarrow
SU(3)_c\otimes SU(2)_L\otimes U(1)_Y\longrightarrow SU(3)_c\otimes
U(1)_Q\] and at the same time give masses to the fermion fields in
our model. With this in mind let us introduce the following set of
Higgs scalars:\\ $\phi_1=(1,\bar{3},-1/3)$ with Vacuum Expectation
Value (VEV) $\langle\phi_1\rangle^T=(0,0,V)$;
$\phi_2(1,\bar{3},-1/3)$ with VEV
$\langle\phi_2\rangle^T=(0,v/\sqrt{2},0)$, and
$\phi_3(1,\bar{3},2/3)$ with VEV
$\langle\phi_3\rangle^T=(v'/\sqrt{2},0,0)$, with the hierarchy
$V>v\sim v'\sim 250$ GeV, the electroweak breaking scale. The
scale of $V$ could be fixed phenomenologically.

At first glance it looks like only two Higgs triplets are necessary for
the symmetry breaking but, as can be seen, they are not enough to
reproduce a realistic fermion mass spectrum in the model.

\section{The gauge boson sector}
There are a total of 17 gauge bosons in our model. One gauge field
$B^\mu$ associated with $U(1)_X$, the 8 gluon fields associated
with $SU(3)_c$ which remain massless, and another 8 associated
with $SU(3)_L$ and that we write for convenience in the following
way:
\[{1\over 2}\lambda_\alpha A^\mu_\alpha={1\over \sqrt{2}}\left(
\begin{array}{ccc}D^\mu_1 & W^{+\mu} & K^{+\mu} \\ W^{-\mu} &
D^\mu_2 &  K^{0\mu} \\
K^{-\mu} & \bar{K}^{0\mu} & D^\mu_3 \end{array}\right) \]
where $D^\mu_1=A_3^\mu/\sqrt{2}+A_8^\mu/\sqrt{6},\;
D^\mu_2=-A_3^\mu/\sqrt{2}+A_8^\mu/\sqrt{6}$,
and $D^\mu_3=-2A_8^\mu/\sqrt{6}$. $\lambda_i, \; i=1,2,...,8$ are the eight
Gell-Mann matrices normalized as $Tr(\lambda_i\lambda_j)=2\delta_{ij}$,
which allows us to write the charge operator as
\[Q=\frac{\lambda_3}{2}+\frac{\lambda_8}{2\sqrt{3}}+XI_3\]
where $I_3$ is the $3\times 3$ unit matrix.

After breaking the symmetry with $\langle\phi_i\rangle ,\;
i=1,2,3$, and using for the covariant derivative for triplets
$D^\mu=\partial^\mu-i{g\over 2} \lambda_\alpha
A^\mu_\alpha-ig'XB^\mu$, we get the following mass terms for the
charged gauge bosons of the electroweak sector:
$M^2_{W^\pm}={g^2\over 4}(v^2+v'^2), \; M^2_{K^\pm}={g^2\over
4}(2V^2+v'^2), \; M^2_{K^0(\bar{K}^0)}={g^2\over 4}(2V^2+v^2)$.
For the three neutral gauge bosons we get a mass term of the form:
\[M=V^2(\frac{g'B^\mu}{3}-\frac{gA_8^\mu}{\sqrt{3}})^2 +
\frac{v^2}{8}(\frac{2g'B^\mu}{3}-gA^\mu_3
+\frac{gA_8^\mu}{\sqrt{3}})^2
+\frac{v'^2}{8}(gA_3^\mu-\frac{4g'B^\mu}{3}+\frac{gA^\mu_8}{\sqrt{3}})^2\]

By diagonalizing $M$ we get the physical neutral gauge bosons which are
defined through the mixing angle $\theta$ and $Z_\mu,\; Z'_\mu$ by:
\begin{eqnarray}\nonumber
Z_1^\mu&=&Z_\mu \cos\theta+Z'_\mu \sin\theta \; ,\\ \nonumber
Z_2^\mu&=&-Z_\mu \sin\theta+Z'_\mu \cos\theta \; ,\\ \nonumber
\tan(2\theta)&=&\frac{-\sqrt{12}C_W(1-T_W^2/3)^{1/2}[v'^2(1+T_W^2)-v^2(1-T_W^2)]}
{3(1-T_W^2/3)(v^2+v'^2)-C_W^2[8V^2+v^2(1-T_W^2)^2+v'^2(1+T_W^2)^2]}
\; ,
\end{eqnarray}
where the photon field $A^\mu$, and the fields $Z_\mu$ and $Z'_\mu$ are given by
\begin{eqnarray}\label{foton}
A^\mu&=&S_W A_3^\mu + C_W\left[\frac{T_W}{\sqrt{3}}A_8^\mu+
(1-T_W^2/3)^{1/2}B^\mu\right]\; ,  \\ \label{z} Z^\mu&=& C_W
A_3^\mu - S_W\left[\frac{T_W}{\sqrt{3}}A_8^\mu+
(1-T_W^2/3)^{1/2}B^\mu\right] \; ,\\ \label{zp}
Z'^\mu&=&-(1-T_W^2/3)^{1/2}A_8^\mu+\frac{T_W}{\sqrt{3}}B^\mu  \; ,
\end{eqnarray}
$S_W=\sqrt{3}g'/\sqrt{3g^2+4g'^2}$ and $C_W$ are the sine and
cosine of the electroweak mixing angle respectively and
$T_W=S_W/C_W$. Also we can identify the $Y$ hypercharge associated
with the SM gauge boson as:
\begin{equation}\label{y}
Y^\mu=\left[\frac{T_W}{\sqrt{3}}A_8^\mu+
(1-T_W^2/3)^{1/2}B^\mu\right].
\end{equation}
In the limit $\sin\theta\rightarrow 0$, $M_{Z}=M_{W^{\pm}}/C_W$,
this limit is obtained either by demanding $V\rightarrow\infty$ or
$v'^2=v^2(C_W^2-S_W^2)\equiv v^2C_{2W}$.

\subsection{Charged currents}
The interactions among the charged vector fields with leptons are
\begin{eqnarray}\nonumber
H^{CC}&=&{g\over \sqrt{2}}[W^+_\mu(\bar{u}_L\gamma^\mu d_L-
\bar{\nu}_{eL}\gamma^\mu e^-_L-\bar{N}^0_{2L}\gamma^\mu E^-_L-
\bar{E}^+_L\gamma^\mu N^0_{4L}) \\ \nonumber & &
+K^+_\mu(\bar{u}_L\gamma^\mu D_L- \bar{N}^0_{1L}\gamma^\mu
e^-_L-\bar{N}^0_{3L}\gamma^\mu E^-_L- \bar{e}^+_L\gamma^\mu
N^0_{4L}) \\ & & +K^0_\mu(\bar{d}_L\gamma^\mu D_L-
\bar{N}^0_{1L}\gamma^\mu \nu_{eL}-\bar{N}^0_{3L}\gamma^\mu
N^0_{2L}- \bar{e}^+_L\gamma^\mu E^+_L)] + h.c.,
\end{eqnarray}
which implies that the interactions with the $K^\pm$ and
$K^0(\bar{K}^0)$ bosons violate the lepton number and the weak
isospin. Notice also that the first two terms in the previous
expression constitute the charged weak current of the SM as far as
we identify $W^\pm$ as the $SU(2)_L$ charged left-handed weak
bosons.

\subsection{Neutral currents}
The neutral currents $J_\mu(EM),\; J_\mu(Z)$ and $J_\mu(Z')$
associated with the Hamiltonian $H^0=eA^\mu J_\mu(EM)+{g\over
{C_W}}Z^\mu J_\mu(Z) + {g'\over \sqrt{3}}Z'^\mu J_\mu(Z')$ are:
\begin{eqnarray}\nonumber
J_\mu(EM)&=&{2\over 3}\bar{u}\gamma_\mu u-{1\over
3}\bar{d}\gamma_\mu d -{1\over 3}\bar{D}\gamma_\mu D-
\bar{e}^-\gamma_\mu e^-- \bar{E}^-\gamma_\mu E^-   \\ \nonumber
&=&\sum_f q_f\bar{f}\gamma_\mu f \\ \nonumber
J_\mu(Z)&=&J_{\mu,L}(Z)-S^2_WJ_\mu(EM)\\
J_\mu(Z')&=&T_WJ_\mu(EM)-J_{\mu,L}(Z')
\end{eqnarray}
where $e=gS_W=g'C_W\sqrt{1-T_W ^2/3}>0$, is the electric charge, $q_f$
is the electric charge of the fermion $f$ in units of $e$, $J_\mu(EM)$
is the electromagnetic current (vectorlike as it should be), and the
left-handed currents are
\begin{eqnarray} \nonumber
J_{\mu,L}(Z)&=&{1\over 2}(\bar{u}_L\gamma_\mu u_L-
\bar{d}_L\gamma_\mu d_L+\bar{\nu}_{eL}\gamma_\mu \nu_{eL}-
\bar{e}^-_L\gamma_\mu e^-_L+\bar{N}^0_{2}\gamma_\mu N^0_{2}
-\bar{E}^-\gamma_\mu E^-) \\ \nonumber &=&\sum_f
T_{3f}\bar{f}_L\gamma_\mu f_L \\ \nonumber
J_{\mu,L}(Z')&=&S_{2W}^{-1}(\bar{u}_L\gamma_\mu u_L-
\bar{e}^-_L\gamma_\mu e^-_L-\bar{E}^-_L\gamma_\mu E^-_L-
\bar{N}^0_{4L}\gamma_\mu N^0_{4L})\\ \nonumber &+&
T_{2W}^{-1}(\bar{d}_L\gamma_\mu d_L-\bar{E}^+_L\gamma_\mu E^+_L -
\bar{\nu}_{eL}\gamma_\mu \nu_{eL} -\bar{N}^0_{2L}\gamma_\mu
N^0_{2L}) \\ \nonumber &-& T_W^{-1}(\bar{D}_L\gamma_\mu D_L -
\bar{e}^+_L\gamma_\mu e^+_L - \bar{N}^0_{1L}\gamma_\mu
N^0_{1L}-\bar{N}^0_{3L}\gamma_\mu N^0_{3L}) \\ &=&\sum_f
T_{9f}\bar{f}_L\gamma_\mu f_L
\end{eqnarray}
where $S_{2W}=2S_WC_W,\; T_{2W}=S_{2W}/C_{2W}, \;
\bar{N}^0_{2}\gamma_\mu N^0_{2}=\bar{N}^0_{2L}\gamma_\mu N^0_{2L}
+\bar{N}^0_{2R}\gamma_\mu N^0_{2R}= \bar{N}^0_{2L}\gamma_\mu
N^0_{2L} -\bar{N}^{0c}_{2L}\gamma_\mu N^{0c}_{2L}=
\bar{N}^0_{2L}\gamma_\mu N^0_{2L} -\bar{N}^0_{4L}\gamma_\mu
N^0_{4L}$, and similarly $\bar{E}\gamma_\mu E=\bar{E}^-_L\gamma_\mu
E^-_L - \bar{E}^+_L\gamma_\mu E^+_L$. In this way
$T_{3f}=Dg(1/2,-1/2,0)$ is the third component of the weak isospin
acting on the representation 3 of $SU(3)_L$ (the negative when
acting on $\bar{3}$), and $T_{9f}=Dg(S_{2W}^{-1}, T_{2W}^{-1},
-T_W^{-1})$ is a convenient $3\times 3$ diagonal matrix acting on
the representation 3 of $SU(3)_L$ (the negative when acting on
$\bar{3}$). Notice that $J_\mu(Z)$ is just the generalization of
the neutral current present in the SM, which allows us to identify
$Z_\mu$ as the neutral gauge boson of the SM, which is consistent
with Eqs.(\ref{foton}), (\ref{z}) and (\ref{y}).

The couplings of the physical states $Z_1^\mu$ and $Z_2^\mu$ are then
given by:
\begin{eqnarray} \nonumber
H^{NC}&=&\frac{g}{2C_W}\sum_{i=1}^2Z_i^\mu\sum_f\{\bar{f}\gamma_\mu
[a_{iL}(f)(1-\gamma_5)+a_{iR}(f)(1+\gamma_5)]f\} \\
      &=&\frac{g}{2C_W}\sum_{i=1}^2Z_i^\mu\sum_f\{\bar{f}\gamma_\mu
      [g(f)_{iV}-g(f)_{iA}\gamma_5]f\}
\end{eqnarray}
where
\begin{eqnarray}
a_{1L}(f)&=&\cos\theta(T_{3f}-q_fS^2_W)-\frac{g'\sin\theta
C_W}{g\sqrt{3}} (T_{9f}-q_fT_W)\;, \\ \nonumber
a_{1R}(f)&=&-q_fS_W(\cos\theta
S_W-\frac{g'\sin\theta}{g\sqrt{3}})\;,\\ \nonumber
a_{2L}(f)&=&-\sin\theta(T_{3f}-q_fS^2_W)-\frac{g'\cos\theta
C_W}{g\sqrt{3}} (T_{9f}-q_fT_W)\;, \\ \nonumber
a_{2R}(f)&=&q_fS_W(\sin\theta S_W+\frac{g'\cos\theta}{g\sqrt{3}})
\end{eqnarray}
and
\begin{eqnarray}
g(f)_{iV}&=&a(f)_{iL}+a(f)_{iR}\;, \\ \nonumber
g(f)_{iA}&=&a(f)_{iL}-a(f)_{iR}\; ,
\end{eqnarray}
so, when the algebra gets done we get:
\begin{eqnarray} \nonumber
g(f)_{1V}&=&\cos\theta(T_{3f}-2S_W^2q_f)-\frac{g'\sin\theta}{g\sqrt{3}}
(T_{9f}C_W-2q_fS_W)\;, \\ \nonumber
g(f)_{2V}&=&-\sin\theta(T_{3f}-2S_W^2q_f)-\frac{g'\cos\theta}{g\sqrt{3}}
(T_{9f}C_W-2q_fS_W) \;,\\ \nonumber 
g(f)_{1A}&=&\cos\theta
T_{3f}-\frac{g'\sin\theta}{g\sqrt{3}}T_{9f}C_W\;, \\ \nonumber
g(f)_{2A}&=&-\sin\theta
T_{3f}-\frac{g'\cos\theta}{g\sqrt{3}}T_{9f}C_W\;,
\end{eqnarray}
to be compared with the SM values $g(f)_{1V}^{SM}=T_{3f}-2S_Wq_f$
and $g(f)_{1A}^{SM}=T_{3f}$. The values of $g_{iV},\; g_{iA}$ with
$i=1,2$ are listed in Tables I and II. As we can see, in the limit
$\theta=0$ the couplings of $Z_1^\mu$ to the ordinary leptons and
quarks are the same as in the SM, and due to this we can test the new
physics beyond the SM.


\begin{center}
TABLE I. The $Z_1^\mu\longrightarrow \bar{f}f$ couplings.
\begin{tabular}{||l||c|c||}\hline\hline
f& $g(f)_{1V}$ & $g(f)_{1A}$ \\ \hline\hline u& $({1\over
2}-{4S_W^2\over 3})[\cos\theta-\sin\theta /(4C_W^2-1)^{1/2}]$ &
${\cos\theta\over 2}-\sin\theta/[2(4C_w^2-1)^{1/2}]$ \\ \hline d&
$\cos\theta (-{1\over 2}+{2S_W^2 \over 3})-
\frac{\sin\theta}{(4C_W^2-1)^{1/2}}({1\over 2} - {S_W^2\over 3})$
& $-{1\over 2}\{\cos\theta + \sin\theta
C_{2W}/[2(4C_W^2-1)^{1/2}]\}$ \\ \hline D& ${2S_W^2\cos\theta
\over 3}+\sin\theta (1-{5\over 3}S_W^2)/(4C_W^2-1)^{1/2}$ &
$C_W^2\sin\theta /(4C_W^2-1)^{1/2}$ \\ \hline $e^-$& $\cos\theta
(-{1\over 2}+2S_W^2)+ \frac{3\sin\theta}{(4C_W^2-1)^{1/2}}({1\over
2} - S_W^2)$ & $ -{\cos\theta\over
2}+\frac{\sin\theta}{(4C_W^2-1)^{1/2}}({1\over 2}-C_W^2)$
 \\ \hline
$E^-$& $\cos\theta (-1+2S_W^2)-
\frac{S_W^2\sin\theta}{(4C_W^2-1)^{1/2}}$ &
$C_W^2\sin\theta /(4C_W^2-1)^{1/2}$ \\ \hline
$\nu_e,\; N_2^0$ &
${1\over 2}[\cos\theta+\sin\theta(1-2S_W^2)/(4C_W^2-1)^{1/2}] $ &
${1\over 2}(\cos\theta+\sin\theta(1-2S_W^2)/(4C_W^2-1)^{1/2} $ \\ \hline
$N_1^0,\; N_3^0$ &
$-C_W^2\sin\theta /(4C_W^2-1)^{1/2}$& $-C_W^2\sin\theta /(4C_W^2-1)^{1/2}$
\\ \hline
$N_4^0$ & $-{1\over 2}[\cos\theta-\sin\theta/(4C_W^2-1)^{1/2}]$ &
$-{1\over 2}[\cos\theta-\sin\theta/(4C_W^2-1)^{1/2}]$ \\ \hline\hline
\end{tabular}
\end{center}

\vspace{2cm}

\begin{center}
TABLE II. The $Z_2^\mu\longrightarrow \bar{f}f$ couplings.
\begin{tabular}{||l||c|c||}\hline\hline
f& $g(f)_{2V}$ & $g(f)_{2A}$ \\ \hline\hline u& $({1\over
2}-{4S_W^2\over 3})[-\sin\theta-\cos\theta /(4C_W^2-1)^{1/2}]$&
${-\sin\theta\over 2}-\cos\theta/[2(4C_w^2-1)^{1/2}]$ \\ \hline d&
$-\sin\theta (-{1\over 2}+{2S_W^2 \over 3})-
\frac{\cos\theta}{(4C_W^2-1)^{1/2}}({1\over 2} - {S_W^2\over 3})$
& $-{1\over 2}\{-\sin\theta + \cos\theta
C_{2W}/[2(4C_W^2-1)^{1/2}]\}$ \\ \hline D& ${-2S_W^2\sin\theta
\over 3}+\cos\theta (1-{5\over 3}S_W^2)/(4C_W^2-1)^{1/2}$ &
$C_W^2\cos\theta /(4C_W^2-1)^{1/2}$ \\ \hline $e^-$& $-\sin\theta
(-{1\over 2}+2S_W^2)+ \frac{3\cos\theta}{(4C_W^2-1)^{1/2}}({1\over
2} - S_W^2)$ & $ {\sin\theta\over
2}+\frac{\cos\theta}{(4C_W^2-1)^{1/2}}({1\over 2}-C_W^2)$
 \\ \hline
$E^-$& $-\sin\theta (-1+2S_W^2)-
\frac{S_W^2\cos\theta}{(4C_W^2-1)^{1/2}}$ &
$C_W^2\cos\theta /(4C_W^2-1)^{1/2}$ \\ \hline
$\nu_e,\; N_2^0$ &
${1\over 2}[-\sin\theta+\cos\theta(1-2S_W^2)/(4C_W^2-1)^{1/2}] $ &
${1\over 2}(-\sin\theta+\cos\theta(1-2S_W^2)/(4C_W^2-1)^{1/2} $ \\ \hline
$N_1^0,\; N_3^0$ &
$-C_W^2\cos\theta /(4C_W^2-1)^{1/2}$& $-C_W^2\cos\theta /(4C_W^2-1)^{1/2}$
\\ \hline
$N_4^0$ & ${1\over 2}[\sin\theta+\cos\theta/(4C_W^2-1)^{1/2}]$ &
${1\over 2}[\sin\theta+\cos\theta/(4C_W^2-1)^{1/2}]$ \\ \hline\hline
\end{tabular}
\end{center}

\section{Fermion masses}
The Higgs scalars introduced in section 3 not only break the symmetry in an
appropriate way, but produce the folowing mass terms for the fermions of
the model:
\subsection{Quark masses}
For the quark sector we can write the following Yukawa terms:
\begin{equation}\label{yuq}
{\cal L}_Y^Q=Q_L^TC(h_u\phi_3u_L^c+h_{D}\phi_1D_L^c+h_{d}\phi_2d_L^c
+h_{dD}\phi_2D_L^c+h_{Dd}\phi_1d_L^c) + h.c.
\end{equation}
where $h_\eta, \; \eta = u, D, d, dD, Dd$, are Yukawa
couplings of order one and $C$ is the charge conjugate operator.
>From Eq.(\ref{yuq}) we get for the up quark sector a mass term
$m_u=h_uv'/\sqrt{2}$, and for the down quark sector a mass matrix
in the basis $(d,D)_L$ of the form:
\begin{equation}
M_{dD}=\left(\begin{array}{cc}
h_{d}v/\sqrt{2} & h_{dD}v/\sqrt{2} \\
h_{Dd}V & h_{D}V \end{array} \right).
\end{equation}
For the particular case $h_{d}=h_{dD}=h_{Dd}=h_{D}\equiv h$, the
mass eigenvalues of the previous matrix are $m_d=0$ and
$m_D=h(V+v/\sqrt{2})$. Since there is not a physical reason for
the Yukawas to be equal, let us calculate the mass eigenvalues as
a function of $|M_{dD}|\equiv (h_dh_D-h_{dD}h_{Dd})$ the
determinant of the Yukawas, and in the expansion $v/V$. Then the
algebra shows that $m_D\simeq
h_{D}V+h_dv/\sqrt{2}-|M_{dD}|v^2/\sqrt{2}Vh_{D}+...$ and
$m_d\simeq
v|M_{dD}|(1+h_{dD}h_{Dd}v/\sqrt{2}Vh_{D}^3+...)/\sqrt{2}h_D$. This
expansion can be used to explain the experimental values for the
quark masses in any one of the three families (for the third
family for example we may choose the particular values
$h_b=h_B=0.8, \; h_{bB}=0.9$ and $h_{Bb}=0.7$ in order to get
$m_b\sim 10^{-2}m_t$, etc.). Since the model does not allow us to
calculate the Yukawa couplings, the fermion masses can not be
predicted in this model. But, contrary to what happen in the SM,
in the context of this model we can implement, in a reasonable way,
the weak isospin breaking in the quark sector for the three
families (what we have here is a particular realization of the
general analysis presented in Ref.\cite{rosner}).

\subsection{Lepton masses}
For the lepton sector we can write the following Yukawa terms:
\begin{equation}
{\cal L}_Y^l=\epsilon_{abc}[\psi_L^aC(h_1\phi_L^b\phi_3^c+h_2\chi_L^b\phi_1^c
+h_3\chi_L^b\phi_2^c)+\phi_L^aC(h_4\chi_L^b\phi_1^c+h_5\chi_L^b\phi_2^c)]+h.c..
\end{equation}
where $a,b,c$ are $SU(3)_L$ tensor indices and the Yukawas are
again of order one. These new terms produce in the basis (e,E) the
mass matrix
\begin{equation}
M_{eE}=\left(\begin{array}{cc} -h_{3}v/\sqrt{2} & -h_{5}v/\sqrt{2}
\\ h_{2}V & h_{4}V \end{array} \right)
\end{equation}
with eigenvalues
$m_E=h_{4}V-h_3v/\sqrt{2}-|M_{eE}|v^2/\sqrt{2}Vh_{4}+...$ and
$m_e\simeq
v|M_{eE}|(1-h_{2}h_{5}v/\sqrt{2}Vh_{4}^3+...)/\sqrt{2}h_4$, where
$|M_{eE}|=h_2h_5-h_3h_4$, with similar consequences as in the down
quark sector.

For the neutral leptons in the basis $(\nu_e,N_1,N_2.N_3,N_4)$ we
get the mass matrix
\begin{equation}
M_N=\left(\begin{array}{ccccc}
0 & 0 & 0 & h_1v'/\sqrt{2} & -h_2V \\
0 & 0 & -h_1v'/\sqrt{2} & 0 & h_3v/\sqrt{2} \\
0 & -h_1v'/\sqrt{2} & 0 & 0 & -h_4V \\
h_1v'/\sqrt{2} & 0 & 0 & 0 & h_5v/\sqrt{2} \\
-h_2V & h_3v/\sqrt{2} & -h_4V & h_5v/\sqrt{2} & 0 \end{array}\right).
\end{equation}
For the particular case $h_2=h_4=h$ and $h_3=h_5=h'$ the mass
eigenvalues are $0$, $\pm h_1v'/\sqrt{2}$, $
\pm\sqrt{h_1^2v'^2/2+h'^2v_2^2+2h^2V^2}$ which means that we have
a Majorana neutrino of zero mass and two Dirac neutral particles
with masses $h_1v'/\sqrt{2}$ and
$(h_1^2v'^2/2+h'^2v_2^2+2h^2V^2)^{1/2}$. The eigenvector
associated with the zero mass Majorana neutrino is $\alpha
(-vh',-\sqrt{2}Vh,vh',\sqrt{2}Vh,v'h_1)$ where
$\alpha=(2v^2/h'^2+4V^2h^2+v'^2h_1^2)^{-1/2}$ is a normalization
factor. This amazing result implies, in the context of this model,
that not only the known tiny mass neutral fermion must be a
Majorana particle, but that it is not the electron neutrino but a
mixture of five neutral particles with only one of them associated
with the vertex $\bar{\nu}_eeW$. 

But $h_2h_5\neq h_3h_4$, otherwise we will have a zero mass
electron. To diagonalize $M_N$ for the most general case
($h_i\simeq 1, i=1,2,....,5$ but not equal) is not a simple task.
A tedious algebra shows that, for the most general case, the five
mass eigenvalues are $\pm h_1v'/\sqrt{2}$ (exact values), one
small see-saw value of order $vv'/V$ and two very large values of
order $(\pm V+\eta)$ where $\eta$ is a small see-saw quotient,
which means that, in the context of this model, the five neutral
leptons in each family split as: a Dirac neutrino with a mass of
the order of the electroweak breaking mass scale, a {\it
pseudo-Dirac}\cite{wolfe} neutrino with a very large mass, and a
see-saw majorana neutrino. In the appendix we carry out the
detailed calculation for the particular case $h_2=h_4\neq h_1\neq
h_3\neq h_5$, in the expansion $vv'/V$. The five mass eigenvalues
for that particular case are: $\pm h_1v'/\sqrt{2}$, $\pm
[(\sqrt{2}h_4V+[(h_3+h_5)^2v^2+2(h_1v'\mp (h_3-h_5)v/\sqrt{2})^2]/
(8\sqrt{2}h_4V)]$ and $m_\nu=(h_5-h_3)vv'/2h_4V$. Notice
that $m_\nu$ is suppressed not only for the see-saw quotient
$vv'/V$ but also for the small difference of the Yukawas
$(h_5-h_3)/2h_4$, which implies that $V$ can be lower than $10^{11}$ GeV, the standard seesaw mechanism value.

\section{Constrains on the $(Z^\mu-Z'^{\mu})$ mixing angle and the $Z^{\mu}_2$
mass} 

To bound $\sin\theta$ and $M_{Z_2}$ we use parameters
measured at the $Z$ pole from CERN $e^+e^-$ collier (LEP), SLAC Linear Collider (SLC) and atomic parity violation 
constraints which are given in the table III \cite{lang2001,data}.
The expression for the partial width for $Z^{\mu}_1\rightarrow
f\bar{f}$ is
\begin{equation}
\Gamma(Z^{\mu}_1\rightarrow f\bar{f})=\frac{N_C G_F
M_{Z_1}^3}{6\pi\sqrt{2}}\rho \left[\frac{3\beta-\beta^3}{2}
(g(f)_{1V})^2+\beta^3 (g(f)_{1A})^2\right](1+\delta_f)R_{QCD+QED}
, \label{ancho}
\end{equation}
\noindent where $f$ is an ordinary SM fermion, $Z^{\mu}_1$ is the
physical gauge boson observed at LEP, $N_C=3$ is the number of
colors, $R_{QCD+QED}$ are the QCD and QED corrections and
$\beta=\sqrt{1-4 m_b^2/M_{Z_1}^2}$ is a kinematic factor which is
$1$ for all the SM fermions except for the bottom quark for which
$m_b=5.0$ GeV \cite{hollik}. The factor $\delta_f$ is zero for all
fermions except for the bottom quark for which the contribution
coming from the top quark in the one loop vertex radiative
correction is parametrized as $\delta_b\approx 10^{-2} (-m_t^2/(2
M_{Z_1}^2)+1/5)$ \cite{pich}. The $\rho$ parameter has two
contributions, one is the oblique correction given by
$\delta\rho\approx 3G_F m_t^2/(8\pi^2\sqrt{2})$ and the other is the
tree level contribution to the $(Z_{\mu} - Z'_{\mu})$ mixing, which
can be written as $\delta\rho_V\approx
(M_{Z_2}^2/M_{Z_1}^2-1)\sin^2\theta$. Finally, $g(f)_{1V}$ and
$g(f)_{1A}$ are listed in table I. Using Eq. (\ref{ancho}), $g(f)_{iV}$ and
$g(f)_{iA}$ we can write expressions for LEP and SLC parameters, 
where we are using the following values\cite{data}:
$m_t=174.3$ GeV, $\alpha_s(m_Z)=0.1192$,
$\alpha(m_Z)^{-1}=127.938$, and $S^2_W=0.2333$.

The efective weak charge $Q_W$ in atomic parity violation is
given by
\begin{equation}
Q_W=-2\left[(2Z+N)c_{1u}+(Z+2N)c_{1d}\right],
\end{equation}
where $c_{1q}=2g(e)_{1A}g(q)_{1V}$, $Z$ is the number of protons
and $N$ is the number of neutrons in the atomic nucleus. The $Q_W$
determination for $^{133}_{55} C_s$ has been improved recently
\cite{bennett} to the value
\begin{equation}
Q_W(^{133}_{55} C_s)=-72.06\pm 0.28\pm 0.34,
\end{equation}
result to be compared with the theoretical value for $Q_W$ which
can be written as \cite{marciano}
\begin{equation}
Q_W(^{133}_{55} C_s)=-73.09\pm 0.04 +\Delta Q_W,
\end{equation}
where $\Delta Q_W$ includes the contributions of
new physics. The discrepancy between the SM and the experimental
data is given by \cite{todos}
\begin{eqnarray}
\Delta Q_W&=&Q_W^{exp}-Q_W^{SM}\nonumber \\ &=&1.03\pm 0.44\; .
\end{eqnarray}
Regarding to the $(Z_{\mu}-Z'_{\mu})$ mixing between two neutral
currents, $\Delta Q_W$ can be written as \cite{durkin}
\begin{equation}
\Delta Q_W = \left[\left(1+4\frac{S_W^2}{1-2S_W^2}\right)-Z\right] 
\delta\rho_V +\Delta Q_W',
\end{equation}
where $\Delta Q_W'$ is model dependent and it can be obtained for
our model by using $g(e)_{1A}$ and $g(q)_{1V}$ from Table I. The
analysis for our model gives
\begin{equation}
\Delta Q_W'=(-3.94 Z -6.40 N) \sin\theta + (1.49 Z + 1.81 N)
\frac{M^2_{Z_1}}{M^2_{Z_2}}\; .
\end{equation}

With the expressions for Z pole observables and $\Delta Q_W$ in terms 
of new physics and using experimental
data from LEP and SLC \cite{lang2001} and atomic parity
violation\cite{bennett} as in table III,  we do a $\chi^2$ fit and we find the
best allowed region for $\sin\theta$ vs. $M_{Z_2}$ at 
95$\%$ CL.

\vspace{.5cm}
\begin{center} TABLE III. Experimental data and SM values for the parameters
\begin{tabular}{||l||l|l||} \hline\hline
& Experimental results & SM \\ \hline
$\Gamma_Z$ (GeV) & $2.4952\pm 0.0023$ & $2.4963\pm 0.0016$  \\ \hline
$\Gamma(had)$ (GeV) & $1.7444\pm 0.0020$ & $1.7427\pm 0.0015$  \\ \hline
$\Gamma(l^+l^-)$ (MeV) & $83.984\pm 0.086$ & $84.018\pm 0.028$ \\ \hline
$R_e$ & $20.804\pm 0.050$ & $20.743\pm 0.018$ \\ \hline
$A_{FB}(e)$ & $0.0145\pm 0.0025$ & $0.0165\pm 0.0003$  \\ \hline
$R_b$ & $0.21653\pm 0.00069$ & $0.21572\pm 0.00015$  \\ \hline 
$R_c$ & $0.1709\pm 0.0034$ & $0.1723\pm 0.0001$ \\ \hline
$A_{FB}(b)$ & $0.0990\pm 0.0020$ & $0.1039\pm 0.0009$ \\ \hline
$A_{FB}(c)$ & $0.0689\pm 0.0035$ & $0.0743\pm 0.0007$ \\ \hline
$A_{FB}(s)$ & $0.0976\pm 0.0114$ & $0.1040\pm 0.0009$ \\ \hline
$A_{b}$ & $0.922\pm 0.023$ & $0.9348\pm 0.0001$ \\ \hline
$A_{c}$ & $0.631\pm 0.026$ & $0.6683\pm 0.0005$ \\ \hline
$A_{s}$ & $0.82\pm 0.13$ & $0.9357\pm 0.0001$ \\ \hline
$A_{e}({\cal P}_{\tau})$ & $0.1498\pm 0.0048$ & $0.1483\pm 0.0012$
\\ \hline        
$Q_W(Cs)$ & $-72.06\pm 0.28\pm 0.34$  & $-73.09\pm 0.04$
\\ \hline\hline
\end{tabular} \end{center} \noindent

In Fig.1 we display the allowed region for $\sin\theta$ vs.
$M_{Z_2}$ at 95$\%$ confidence level. When
$\sin\theta$ goes to zero, the contribution of $\delta\rho_V$ to
$\Gamma(ff)$ decouples, which allow us to get a lower
bound for $M_{Z_2}$ (the bounds come from $\Delta Q_W$). At
95$\%$ CL the allowed region gives
\begin{eqnarray}
-0.0015\leq &\sin\theta& \leq 0.0048  \;, \nonumber \\ 600\; GeV \leq &
M_{Z_2},
\end{eqnarray}
which implies that the mass of the new neutral gauge boson is compatible with
the bound got in $p\bar{p}$ collisions at  Tevatron \cite{fermi}.
A model independent
analysis, similar to the one presented here has been reported by
J. Erler and P. Langacker in Ref.\cite{todos}.

\section{Conclusions}
We have presented an anomaly-free model based on the local gauge
group $SU(3)_c\otimes SU(3)_L\otimes U(1)_X$ which is a subgroup
of the electroweak-strong unification group $E_6$. We break the
gauge symmetry down to $SU(3)_c\otimes U(1)_{Q}$ and at the same
time give masses to all the fermion fields in the model in a
consistent way by using three different Higgs scalars $\phi_i,\:
i=1,2,3$ which set two different mass scales; $v\sim v'\sim 250$
GeV $<<V$. By using experimental results from LEP, SLC and atomic
parity violation we bound the mixing angle and the mass of the
additional neutral current to be $-0.0015\leq\sin\theta<0.0048$
and $600\; GeV\leq M_{Z_2}$  at 95$\%$ CL.

The most outstanding result of our analysis is the mass spectrum
of the fermion fields in the model which arises as a consequence
of the mixing between ordinary fermions and their exotic
counterparts without the need of different scale Yukawa couplings.
Conspicuously, the five neutral leptons in the model split as a
Dirac neutrino with a mass of the order of the electroweak
breaking mass scale $v$, a {\it pseudo-Dirac} neutrino with a
large mass of order $V$ and a tiny see-saw Majorana neutrino which
is not the partner of the lightest charged lepton at the $W^{\pm}$
vertex, but a mixture of the five neutral states in the weak
bases. 

Notice that all the new states are vector-like with respect to the
SM quantum numbers. They consist of an isosinglet quark $D$ of
electric charge $-1/3$, a lepton isodoublet $(E^-,N_2^0)_L^T$ with
$N_{2L}^{0c}=N_{4L}^0$ in the weak basis, and two neutral lepton
isosinglets $N_{1L}^0$ and $N_{3L}^0$. Since $V$ can be of the
order of a few TeV, these exotic fermions could be accessible in
forth coming searches at the Fermilab TeVatron collider or at the
LHC under construction at CERN.

But two neat predictions of this model are: first, the existence
of a Dirac neutrino for each family with a mass of the order of
the electroweak mass scale; second, the existence of a new neutral
gauge boson with a mass just below the TeV scale. The experimental
signature for this particle may be present already at the LEP II
$e^+e^-$ collider data , or just around the corner of the Fermilab
TeVatron output.

\section{ACKNOWLEDGMENTS}
This work was partially supported by BID and Colciencias in Colombia.

\newpage

\begin{center}
{\bf APPENDIX}
\end{center}

In this appendix we diagonalize perturbatively the mass matrix for
the neutral leptons which appears in section {\bf 5.2}. First let
us simplify the notation with the following definitions:
$a=h_1v'/\sqrt{2}, \, B=-h_2V,\;c=h_3v/\sqrt{2},\;b=-h_4V$ and
$C=h_5v/\sqrt{2}$. Then the mass terms for the neutral fermions
now reads:
$M_N=B\nu_eN_4+a\nu_eN_3-aN_1N_2+cN_1N_4+bN_2N_4+CN_3N_4+h.c.$;
which in the basis $(\nu_e,N_1,N_2,N_3,N_4)$ becomes:

\[M_N=\left(\begin{array}{ccccc}
0  & 0  & 0  & a & B \\ 0  &  0  &  -a  &  0  &  c  \\ 0  &  -a &
0  &  0  &  b  \\  a  &  0  &  0  &  0  &  C  \\  B  &  c  &  b  &
C &  0  \end{array}  \right) \]

Making a unitary transformation of $M_N$ to the new basis:

\begin{eqnarray}
E_1&=&\alpha[-(b+c),(B-C),(B-C),(b+c),0] \\ \nonumber
   E_2&=&\beta[(b-c),(B+C),-(B+C),(b-c),0] \\ \nonumber
   E_3&=&\beta[(B+C),-(b-c),(b-c),(B+C),0] \\  \nonumber
   E_4&=&\alpha[(B+C),(b+c),(b+c),-(B-C),0] \\ \nonumber
   E_5&=&e_5=[0,0,0,0,1]  \nonumber
\end{eqnarray}

\noindent where $\alpha=1/[2(b+c)^2+2(B-C)^2]^{1/2}$ and $ \beta
=1/[2(b-c)^2+2(B+C)^2]^{1/2}$, we get the mass matrix in the
following block diagonal form:

\[M´_N=\left(\begin{array}{ccccc}
-a  & 0  & 0  & 0 & 0 \\ 0  &  a  &  0  &  0  &  0  \\ 0  &  0 &
a& 0  &  1/2\beta  \\  0  &  0  &  0  &  -a  &  1/2\alpha  \\ 0  &
0 & 1/2\beta & 1/2\alpha &  0
\end{array}  \right) \].

For the non-diagonal $3\times 3$ block we use the approximation
$B=b>>c\sim C$ and the perturbative expansion
$v^mv´^{n-m}/V^{n-1}$ for $n=1,2,3...$ and $m=0,1,2,...\leq n$.
Then we get:
\[M´_{3N}=\left(\begin{array}{ccc} 0  &  0  &  b  \\
0  &  0  &  b  \\  b  &  b  &  0  \end{array}\right)+
\left(\begin{array}{ccc}a  &  0  &
\frac{(C-c)}{2}+\frac{(C+c)^2}{8b} \\ 0  &  -a
&\frac{-(C-c)}{2}+\frac{(C+c)^2}{8b}\\
\frac{(C-c)}{2}+\frac{(C+c)^2}{8b} &
\frac{-(C-c)}{2}+\frac{(C+c)^2}{8b} & 0 \end{array} \right).\]

\noindent Rotating now $M´_{3N}$ with

\[R=\left(\begin{array}{ccc} 1/\sqrt{2}
& -1/\sqrt{2} & 0 \\ 1/2  &  1/2  &  1/\sqrt{2} \\ 1/2  &  1/2  &
-1/\sqrt{2} \end{array} \right),\]

\noindent we finally get

\[\left(\begin{array}{ccc} 0  &  \frac{a}{\sqrt{2}}+\frac{(C-c)}{2}
&\frac{a}{\sqrt{2}}-\frac{(C-c)}{2} \\
\frac{a}{\sqrt{2}}+\frac{(C-c)}{2} & \sqrt{2}[b +
\frac{(C+c)^2}{8b}] &  0  \\ \frac{a}{\sqrt{2}}-\frac{(C-c)}{2} &
0 &  -\sqrt{2}[b + \frac{(C+c)^2}{8b}] \end{array}\right), \]

\noindent which we diagonalize using matrix perturbation theory up
to second order in the perturbation \cite{landau}. After the
algebra is done we get the following three eigenvalues:
$m_1\simeq\frac{a(c-C)}{b}, \;
m_2\simeq\sqrt{2}b+\frac{(c+C)^2}{4\sqrt{2}b}+\frac{1}{\sqrt{2}b}({a
\over\sqrt{2}}+\frac{(C-c)}{2})^2$ and $m_3 \simeq -
\sqrt{2}b-\frac{(c+C)^2}{4\sqrt{2}b}-\frac{1}{\sqrt{2}b}({a
\over\sqrt{2}}-\frac{(C-c)}{2})^2$. Notice that for $h_3=h_5\;\;
(c=C)$ we get a zero mass Majorana neutrino.

\newpage

\begin{figure}[t]
\epsfxsize=10 truecm \epsfysize=12 truecm \centerline{\epsffile[15 18 550 550]{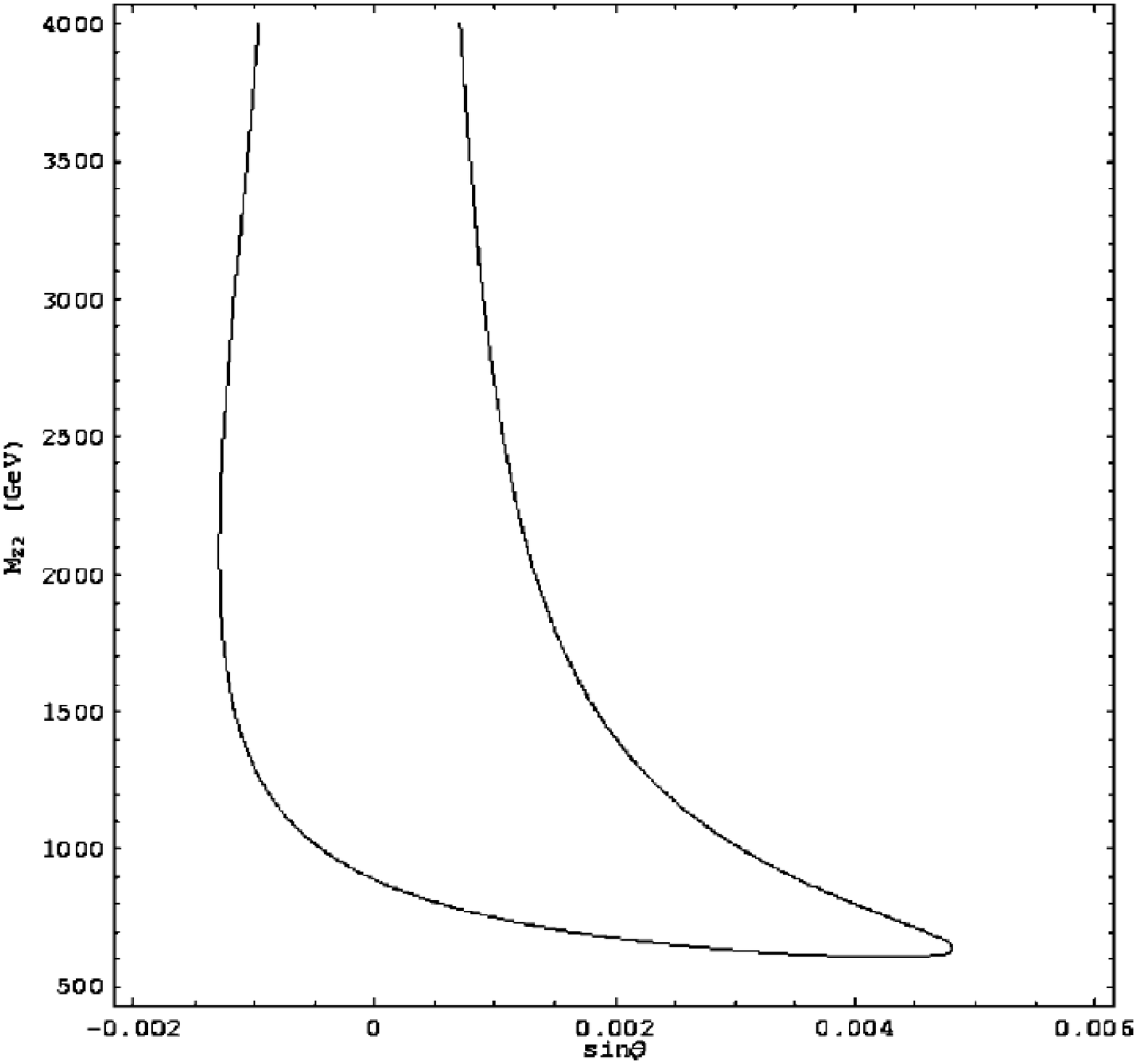}}
\end{figure}

 \noindent Fig.1  Contour plot displaying the allowed region
for $\sin\theta$ vs. $M_{Z_2}$ at 95$\%$ CL.

\end{document}